\documentclass[prl,twocolumn,floatfix,superscriptaddress,citeautoscript]{revtex4}

\usepackage{graphicx}
\usepackage{dcolumn}   
\usepackage{bm}        
\usepackage{amssymb}   
\usepackage{verbatim}
\usepackage{multirow}

\usepackage{amsmath}
\usepackage{amsfonts}

\begin{document}

\title{Trapping a single vortex and reducing quasiparticles in a superconducting resonator}



\author{I. Nsanzineza}
\affiliation{Department of Physics, Syracuse University, Syracuse, NY 13244-1130}
\author{B.L.T. Plourde}
\email[]{bplourde@syr.edu}
\affiliation{Department of Physics, Syracuse University, Syracuse, NY 13244-1130}

\date{\today}

\begin{abstract}
Vortices trapped in thin-film superconducting microwave resonators can have a significant influence on the resonator performance. Using a variable-linewidth geometry for a weakly coupled resonator we are able to observe the effects of a single vortex trapped in the resonator through field cooling. For resonant modes where the vortex is near a current antinode, the presence of even a single vortex leads to a measurable decrease in the quality factor and a dispersive shift of the resonant frequency. For modes with the vortex located at a current node, the presence of the vortex results in no detectable excess loss and, in fact, produces an increase in the quality factor. We attribute this enhancement to a reduction in the density of nonequilibrium quasiparticles in the resonator due to 
their trapping and relaxation near the vortex core.
\end{abstract}

\maketitle

Superconducting thin-film microwave resonators play a critical role in many areas including circuits for quantum information processing \cite{Clarke08,Wallraff04} and photon detectors for astrophysical applications \cite{Day:2003}. Frequently these resonators are operated in environments with a non-negligible magnetic field, perhaps due to insufficient magnetic shielding, magnetism in packaging and connector hardware, or pulsed magnetic fields for controlling circuit parameters.

The response of magnetic flux vortices in such resonators has been studied through field-cooled measurements and related to the vortex viscosity and pinning strength in different superconducting films \cite{Song:2009}. In general, trapped vortices were found to cause a reduction in the resonator quality factor, with the magnitude of the effect scaling with the number of vortices, 
as well as a downwards shift in the resonance frequency. 
Patterned surface pinning \cite{Song:2009b} and other vortex-trapping structures \cite{Bothner:2011,Bothner:2012} have been investigated for minimizing the excess loss contributions from vortices for circuits that require operation in large magnetic fields. These previous experiments have all involved resonators with many trapped vortices. The response of a single vortex in such a microwave circuit has not yet been explored.

\begin{figure}
\centering
\includegraphics[width=3.35in]{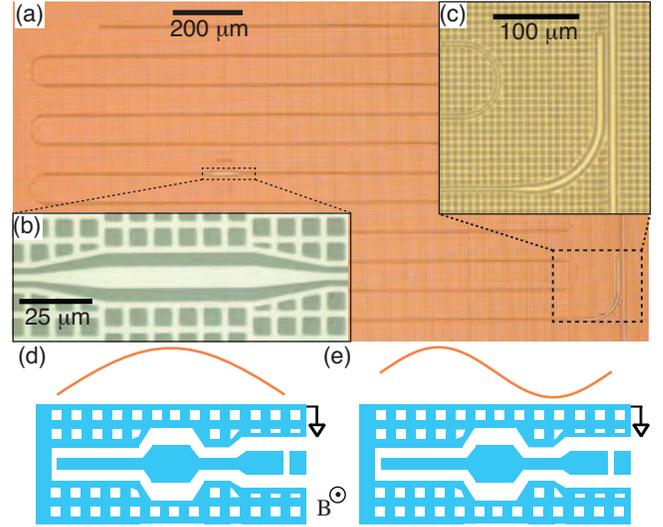}
  \caption{(Color online) Optical micrographs of (a) entire resonator including feed line, (b) close-up of bulge region for vortex trapping near center of resonator, (c) close-up of coupling elbow and feed line. Schematic of resonator without turns (not to scale) along with standing-wave pattern of microwave current for (d) fundamental, (e) first-harmonic resonance. 
\label{fig:setup}}
\end{figure}

In this Letter, we present field-cooled measurements of a coplanar-waveguide resonator with a geometry designed to allow vortex trapping in only a small region over a wide range of magnetic fields. Because the resonator is weakly coupled to the external circuitry and has a reasonably high internal quality factor $Q_i$, we are able to resolve the influence of individual vortices. In addition, we observe a dramatic difference in the effects of the first several trapped vortices on the particular resonance mode that we excite. When the vortices are near an antinode of the current standing-wave pattern, there is a stepwise increase in the loss. However, vortices located near a current node contribute no extra loss, and, in fact, lead to a decrease in the loss, a process that we attribute to enhanced trapping of nonequilibrium quasiparticles due to the cores of the trapped vortices.

In order to control the location of the trapped vortices upon field cooling, we design our resonator to make use of the width dependence of the threshold perpendicular magnetic field for vortex trapping, $B_{th}$. For a trace of width $w$, $B_{th}(w) \sim w^{-2}$, 
although there is also a 
logarithmic correction related to the vortex core energy that can be significant 
\cite{clem98,likharev72}. This relationship has been studied through vortex imaging experiments on superconducting strips of different widths cooled in a range of fields \cite{Stan:2004}
and we have included an analysis of $B_{th}(w)$ for our device in the supplementary material. 
Thus, a wide trace will begin trapping vortices at a smaller field as compared to a narrow trace. Therefore, we design the center conductor of our coplanar-waveguide resonator to be $3\,\mu{\rm m}$ wide over most of its length, with a bulge having a width of $8\,\mu{\rm m}$ for the central $50\,\mu{\rm m}$ along the length of the resonator (Fig.~\ref{fig:setup}). 
Furthermore, the ground plane contains an array of holes that are $5.6\,\mu{\rm m}$ wide and separated by a superconducting web with a linewidth of $2.8\,\mu{\rm m}$ to avoid the trapping of vortices outside of the central bulge region of the center conductor for fields below $B_{th}$ for 
all of the narrower traces on the device.

Our resonator is $17.1\,{\rm mm}$ long and has an elbow-style capacitive coupler to a feed line at one end and an open circuit on the other end. The fundamental resonance corresponds to a half wavelength with 
a current antinode 
at the central bulge. The resonator is patterned from a 60-nm thick Al film on a high-resistivity Si wafer using photolithography followed by a wet-etch process. 

We cool the device on an adiabatic demagnetization refrigerator (ADR) with a $3\,{\rm K}$ pulse-tube cooled stage. The resonator chip is mounted on the cold finger of the ADR and is located at the center of a superconducting Helmholtz coil at $3\,{\rm K}$. We repeatedly heat the cold finger to $\sim 1.5\,{\rm K}$ to exceed $T_c$ for the Al film then cool to $100\,{\rm mK}$ while applying different magnetic fields with the Helmholtz coil. A cryogenic mu-metal can at $3\,{\rm K}$ shields the resonator from stray magnetic fields outside of the cryostat as well as any residual stray fields from the ADR magnet. By cooling in positive and negative magnetic fields applied from the Helmholtz coil and comparing any small asymmetry between measurements of the same vortex-trapping features (not shown), we estimate the component of the background magnetic field perpendicular to our sample to be less than $2\,\mu{\rm T}$.

Upon reaching $100\,{\rm mK}$ for each field-cooling point, we measure the microwave transmission $S_{21}$ through the feed line with a vector network analyzer. 
Following the subtraction of a separate calibration of the magnitude and phase of the background transmission, for each cooling field we fit $S_{21}^{-1}$ in the complex plane with a four-parameter model \cite{Megrant:2012} to extract the total quality factor $Q$ 
for each cooling field. We measure 
$S_{21}$ 
at sufficiently high powers
($\sim 10^5$ photons) 
in order to minimize the loss due to two-level defects on the surfaces and interfaces \cite{Martinis:2005}.

\begin{figure}
\centering
\includegraphics[width=3.35in]{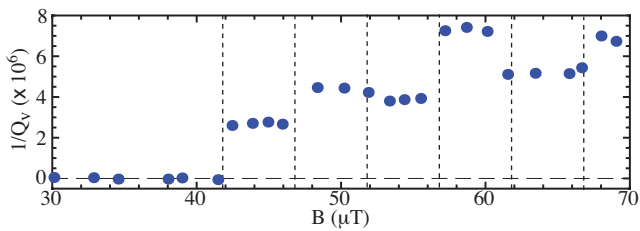}
  \caption{(Color online) $1/Q_v(B)$ for fundamental resonance for cooling fields in 
  vicinity of $B_{th}(8\,\mu{\rm m})$ for 
   central bulge region. Vertical dashed lines correspond to field steps $\Delta B=5\,\mu{\rm T}$.
  \label{fig:fundamental}}
\end{figure}

We observe the fundamental resonance at $3.0713\,{\rm GHz}$ with a coupling quality factor $Q_c=765,000$. For zero-field cooled measurements, we measure $Q=185,000$, thus the resonator is significantly undercoupled 
with internal losses dominating coupling losses $(1/Q=1/Q_i+1/Q_c)$. 
At each cooling field, we extract the loss due to vortices by computing $1/Q_v=1/Q(B)-1/Q(B=0)$ \cite{Song:2009}, thus subtracting out contributions from all other loss mechanisms, such as coupling to external circuitry or dielectric loss. For sufficiently small $B$, we observe $1/Q_v=0$ as there are no vortices trapped in the resonator (Fig. \ref{fig:fundamental}). At a cooling field of $42\,\mu{\rm T}$ there is an abrupt step upwards in $1/Q_v$, which we attribute to the trapping of one vortex in the central bulge. 

The first step in $1/Q_v$ is followed by a series of steps that are spaced by $\Delta B \approx 5\,\mu{\rm T}$. Assuming each step corresponds to an increase in the number of vortices by one, this corresponds to an effective area for vortex trapping of $\Phi_0/\Delta B \approx 400\,\mu{\rm m}^2$, which matches the area of the bulge region in our resonator, where $\Phi_0 \equiv h/2e$ is the magnetic flux quantum. 
While the step widths are quantized, as one would expect for the sequential addition of one vortex, the step heights are clearly not constant, and, in fact, do not always have the same sign, as in the step from $4$ to $5$ vortices. Because $1/Q_v$ depends on the local current density, which will be highly nonuniform across the width of the bulge \cite{Song:2009}, vortices located near the edge of the bulge will contribute more loss compared to a vortex near the centerline. The vortex positions are determined by the random pinning potential in the Al film as well as the intervortex interactions that are present immediately below $T_c$ when the vortices are still mobile, before the vortices become pinned somewhat further below $T_c$ \cite{Bronson:2006}. At our measurement temperature, the superconducting penetration depth is less than $100\,{\rm nm}$ and the vortices 
no longer interact with one another. 
Despite the variations in step height for our measurements, we can estimate an approximate loss per vortex using Eq.~(12) from Ref.~\cite{Song:2009} with 
parameters for the Al film on this device. We obtain a value between $1-5\times10^{-6}$ depending on the vortex location with respect to the current density distribution, consistent with our measured steps in $1/Q_v$.

\begin{figure}
\centering
\includegraphics[width=3.35in]{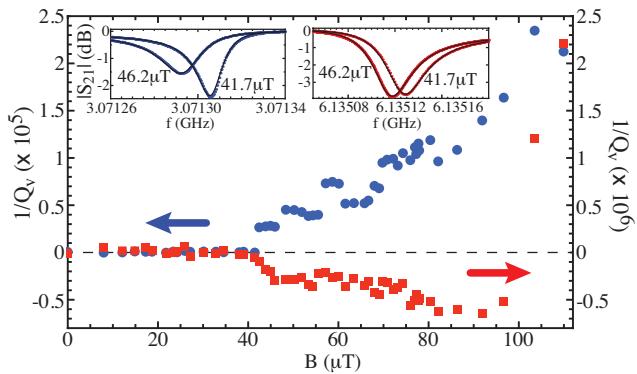}
  \caption{(Color online) $1/Q_v(B)$ for fundamental (blue circles) and first harmonic (red squares) resonance -- note different scales on loss axes. (insets) $|S_{21}(f)|$ for (left) fundamental; (right) harmonic for $B=41.7\,\mu{\rm T}$ (no vortices) and $46.2\,\mu{\rm T}$ (one-vortex step).
  \label{fig:fund-harm}}
\end{figure}

In addition to the fundamental, we can also measure the first harmonic at $6.1351\,{\rm GHz}$, with $Q_c=341,000$, corresponding to a full-wavelength resonance with a current node at the central bulge. Thus, we  expect that vortices trapped in the bulge should contribute no loss to this harmonic resonance, as there is no current present to drive the vortices. However, our measurements of $1/Q_v$ for the harmonic exhibit 
a {\it decrease} to lower loss at the same $B_{th}(8\,\mu{\rm m})$ where we observe the first step upwards in $1/Q_v$ for the fundamental (Fig. \ref{fig:fund-harm}). 
While this downwards trend for the harmonic is clearly visible, it is not as sharp as the upwards step for the fundamental. Because the changes in $1/Q_v$ for the harmonic are about one order of magnitude weaker than those on the fundamental, slight variations in the extracted loss, due perhaps to variations in the temperature of our cryostat or changes in the electromagnetic environment for measuring the resonator from run to run, tend to smooth out small features in $1/Q_v(B)$ for the harmonic. 
$1/Q_v$ continues to decrease for larger cooling fields until a field of $\sim90\,\mu{\rm T}$, at which point there is a significant increase to higher loss values. We attribute this increase at large fields to vortices that begin to trap along the entire length of the resonator for $B>B_{th}(3\,\mu{\rm m})$, where there are significant microwave currents to drive the vortices. 
For $B > 110\,\mu{\rm T}$, the internal losses from the vortices become large enough relative to the coupling loss $1/Q_c$ that we are unable to fit the resonance and extract a value for $Q$. 
%
We have chosen to focus our analysis on the changes in loss for the fundamental and harmonic rather than the shifts in the resonance frequencies. Because the resonator in this experiment is quite narrow over most of its length and thus has a substantial kinetic inductance contribution, nonlinear effects of the superconductor itself dominate the frequency response to changes in the magnetic field, as has been studied previously \cite{Healey:2008}.

We interpret 
the 
decrease in $1/Q_v$ for the harmonic resonance in terms of a reduction in the loss due to quasiparticles $1/Q_{qp}$ 
due to interactions between 
quasiparticles and 
vortex cores. At our measurement temperature of $100\,{\rm mK}$, the density of thermal quasiparticles should be vanishingly small. However, 
several recent investigations 
have demonstrated that without extensive shielding of stray light, superconducting Al circuits measured at millikelvin temperatures can exhibit a significant excess of nonequilibrium quasiparticles with a typical volume density $n_{qp} \sim 10-100\,\mu {\rm m^{-3}}$ \cite{Barends:2011,Corcoles:2011,deVisser:2011}. Blackbody photons emitted by warmer regions of the measurement cryostat, even if only at a few Kelvin, can be sufficiently energetic to break Cooper pairs in Al films due to the relatively small superconducting energy gap. $1/Q_{qp}$ is proportional to the density of quasiparticles in the superconductor $n_{qp}$, thus, this mechanism can lead to excess loss \cite{Martinis:2009}.

Measurements of the effectiveness of different levels of infrared shielding of
Al resonators were reported in Ref.~\cite{Barends:2011} where the cryostat temperature on an ADR was increased while the cold finger was maintained below $150\,{\rm mK}$. With minimal shielding, comparable to our 
setup, the high-power resonator loss was observed to increase with 
cryostat temperature, as one would expect for a blackbody source. We have performed the same measurement on our ADR with an identical resonator to the one presented here after zero-field cooling and observed a similar increase in loss with cryostat temperature (supplemental material). Thus, we conclude that 
nonequilibrium quasiparticles also limit the loss of our resonators at the high power of our measurements. Following the analysis 
and Eq. (1) 
in Ref.~\cite{Barends:2011} and using a kinetic inductance fraction of $0.27$ that we measured on the same cooldown, we obtain $n_{qp} = 50\,\mu{\rm m^{-3}}$ 
in zero field. 

Interactions between quasiparticles and vortices have been studied previously in quasiparticle lifetime experiments \cite{Ullom:1998} and also in the context of tunnel junction photon detectors \cite{Golubov:1993} and NIS coolers \cite{Peltonen:2011}. These all involve many vortices trapped in the superconducting region with the suppressed gap in the vicinity of each vortex core providing a pathway for quasiparticle relaxation and trapping. In Ref.~\cite{Ullom:1998}, quasiparticles were injected with a tunnel junction at one end of an Al strip and their diffusion along the strip was measured with a second tunnel junction some distance away. The quasiparticle flux reaching the detector junction was significantly reduced when a magnetic field was used to generate vortices in the Al strip. This process was modeled with a quasiparticle diffusion equation with an extra recombination term depending on the fraction of nonsuperconducting regions, related to the density of vortices in the film.

We follow a related approach to model 
quasiparticle diffusion in our resonator, but with discrete regions of enhanced recombination localized around each vortex. 
We treat the diffusion process in 1D, neglecting variations in the width of the center conductor of the resonator:
\begin{equation}
D \nabla^2 n_{qp} - \Gamma_R n_{qp}^2 + \gamma_i - \Gamma_v n_{qp} e^{-(x-x_v^i)^2/l_v^2}  = 0.
\label{eq:vortex-eom}
\end{equation}
$D$ is the quasiparticle diffusion constant, which varies with energy, $D(E) = D_n (1-(\Delta/E)^2)^{1/2}$ \cite{Ullom:1998}, where $D_n$ is the normal metal diffusion constant. 
We take $D_n = 60\,{\rm cm^2/s}$ based on previous work on quasiparticle diffusion in Al \cite{Friedrich:1997}. 
$D(E)$ has the strongest variation for quasiparticles with energies just above the gap, $\Delta$, while $D$ only varies by $\sim 15\%$ for energies above $2\Delta$. Because the pair-breaking radiation in our system is likely originating from the $3\,{\rm K}$ shield and warmer portions of our cryostat, the dominant part of this spectrum will lead to the majority of the nonequilibrium  quasiparticles with energies of a few times $\Delta$ and above. Thus, to simplify the analysis while still capturing the essential dynamics, we take $D=D(2\Delta)$. 
We have explored the effects of varying $D$ in our simulations and found that we can obtain reasonable agreement with our measurements over a wide range of $D$ for physically realistic values of the other parameters in the simulations (See supplemental material, which includes Refs.~\cite{Anthore:2003,Pekola:2000}).

$\Gamma_R$ is the effective background quasiparticle recombination rate 
in the Al film and is independent of position. The exact value of $\Gamma_R$ depends on details of phonon trapping and is difficult to obtain precisely. Based on values extracted by others for Al thin films, $\Gamma_R$ can be constrained to $10-100\,\mu{\rm m}^3/s$ \cite{Ullom:1998}. $\gamma_i$ is the quasiparticle generation rate, which we also take to be independent of position, and we adjust the value of $\gamma_i$ in order to match the value of $n_{qp}$ with no vortices present that we obtain from our measured $1/Q_i$ for zero-field cooling.

The final term in Eq.~(\ref{eq:vortex-eom}) represents the quasiparticle-vortex interaction for one vortex centered at $x_v^i$. $\Gamma_v$ corresponds to the 
rate of quasiparticle trapping and relaxation 
in the vicinity of the vortex and thus this term has a strong spatial variation representing the suppression of the gap near the vortex core. We take the spatial dependence to be a Gaussian with a characteristic length-scale $l_v=0.5\,\mu{\rm m}$ based on a treatment in Ref.~\cite{Golubov:1993} of the gap suppression near a vortex using the Usadel equations with a prediction of an effective radius of $\sim 2.7 \xi$. Changing the functional form for this spatial variation or the value of $l_v$ could impact the value of $\Gamma_v$ that we extract, 
but the qualitative outcome would be unchanged. 

We solve Eq.~(\ref{eq:vortex-eom}) with MATLAB using a numerical package involving piecewise Chebyshev polynomial interpolants \cite{Chebfun}. A damped Newton method is applied iteratively with an adaptive mesh 
to deal with the 
micron-scale features in the vicinity of each vortex while solving the nonlinear differential equation over the entire 
%
length $L$ of the resonator. Because the open-ended geometry of our resonator avoids quasiparticle out-diffusion, 
we apply the boundary condition $\partial{n_{qp}}/\partial{x}=0$ at both ends.

\begin{figure}
\centering
\includegraphics[width=3.35in]{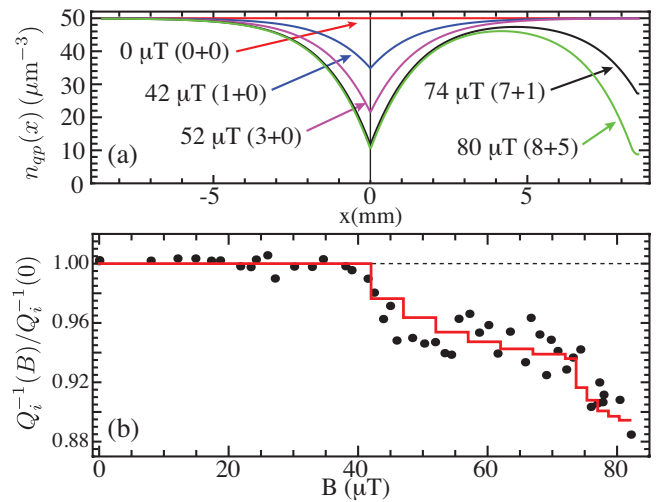}
\caption{(Color online) (a) Simulated $n_{qp}(x)$ for several example
fields. Labels indicate vortex number in 
bulge + 
elbow. (b) Measured $1/Q_i(B)$ for harmonic, normalized by average of $1/Q_i$ below threshold field (points); 
normalized quasiparticle loss on harmonic from simulated $n_{qp}(x)$ 
(solid line).
\label{fig:sims-and-data}}
\end{figure}

We simulate the field dependence of $n_{qp}(x)$ by including one vortex term for each vortex in the distribution for a particular field range. From the analysis of the steps in $1/Q_v$ for the fundamental, we extract the number of trapped vortices for each field range then assign $x_v^i$ for each of these to space them evenly in the middle $50\,\mu{\rm m}$ along $x$, corresponding to the central bulge region. 
We have checked that variations in the exact vortex positions in the bulge region have a negligible impact on our results (supplemental material). 
At a cooling field of $72\,\mu{\rm T}$, following the addition of the sixth vortex to the central bulge, 
there is a more rapid decrease in $1/Q_v$ for the harmonic (Fig.~\ref{fig:fund-harm}). This corresponds to the intermediate $B_{th}(6\,\mu{\rm m})$ for the $6\,\mu{\rm m}$-wide coupling elbow, which is also at a current node. $B_{th}(6\,\mu{\rm m})$ is in between $B_{th}(8\,\mu{\rm m})$ for the bulge and $B_{th}(3\,\mu{\rm m})$ for much of the rest of the resonator. Because the area of the elbow region is about three times larger than that of the central bulge, beyond $72\,\mu{\rm T}$ we add one vortex to the elbow, evenly spaced within the elbow, every $1.7\,\mu{\rm T}$, while continuing to add one vortex to the bulge region every $5\,\mu{\rm T}$.

Figure~\ref{fig:sims-and-data} contains several resulting $n_{qp}(x)$ profiles for four different vortex configurations. 
In order to compare the simulation results with the measured internal loss on the harmonic $1/Q_i(B)$, we 
account for the variation of the standing-wave current along the length of the resonator as described in the supplemental material. 
%
We then compare this with the measured $1/Q_i(B)$ for the harmonic, normalized by the average of $1/Q_i(B)$ for $B<B_{th}(8\,\mu{\rm m})$. 
%
%
We then adjust $\Gamma_v$ for the closest agreement between the simulations and the data. 
We have found that $\Gamma_R=30\,\mu{\rm m}^3/s$, consistent with earlier work for Al films \cite{Ullom:1998}, 
combined with $\Gamma_v=3.5\times10^6\,{\rm s}^{-1}$ provides a good match with the experiment [Fig.~\ref{fig:sims-and-data}(b)], 
although for different $D$ values there are moderately different values of $\Gamma_R$ and $\Gamma_v$ that also provide reasonable agreement with our data (supplemental material).   
The value of $\Gamma_v$ that we extract is in the range of typical electron-phonon scattering rates for Al thin films at low temperatures \cite{Kaplan:1976,Chi:1979}, 
although it is possible that electron-electron scattering in the vicinity of the vortex core may play a role as well \cite{Ullom:1998}. 

While our simulations of $n_{qp}$ provide a reasonable qualitative description of our loss measurements on the harmonic, they do not provide a perfect match to the data. For example, the initial decrease in $1/Q_i$ 
with the first few trapped vortices is not as rapid in our simulations compared to the experiment. In the future, a more sophisticated treatment of the quasiparticle diffusion and interaction with vortices could yield even better agreement and may reveal new features of this interaction.

Future devices could employ patterned pinning sites \cite{Song:2009b} in the trapping region to control the vortex location for further investigations of vortex dynamics and quasiparticle-vortex interactions. The ability to trap vortices in specific regions may be useful in hybrid superconducting-atomic systems as well \cite{Romero:2013}. 

We acknowledge useful discussions with J.M. Martinis, R. McDermott, and C.M. Wilson. 
During the preparation of this work, we became aware of related experiments on vortex trapping 
in superconducting qubits and a similar reduction in quasiparticle density \cite{Wang:2014b,Vool:2014} and we benefited from useful discussions with the authors of these works as well. 
This work was supported by the National Science Foundation under Grant No. DMR-1105197. Device fabrication was performed at the Cornell NanoScale Facility, a member of the National Nanotechnology Infrastructure Network, which is supported by the National Science Foundation (Grant ECS-0335765).

\bibliography{resonator-vortex2}

\widetext
\clearpage
\begin{center}
\textbf{\large Supplementary information to the manuscript ``Trapping a single vortex and reducing quasiparticles in a superconducting resonator"}
\end{center}





\setcounter{equation}{0}
\setcounter{figure}{0}
\setcounter{table}{0}
\setcounter{page}{1}
\makeatletter
\renewcommand{\theequation}{S\arabic{equation}}
\renewcommand{\thefigure}{S\arabic{figure}}
\renewcommand{\bibnumfmt}[1]{[S#1]}
\renewcommand{\citenumfont}[1]{S#1}

\section{configuration of cryostat and measurements}

\begin{figure}[b]
\centering
\includegraphics[width=3.35in]{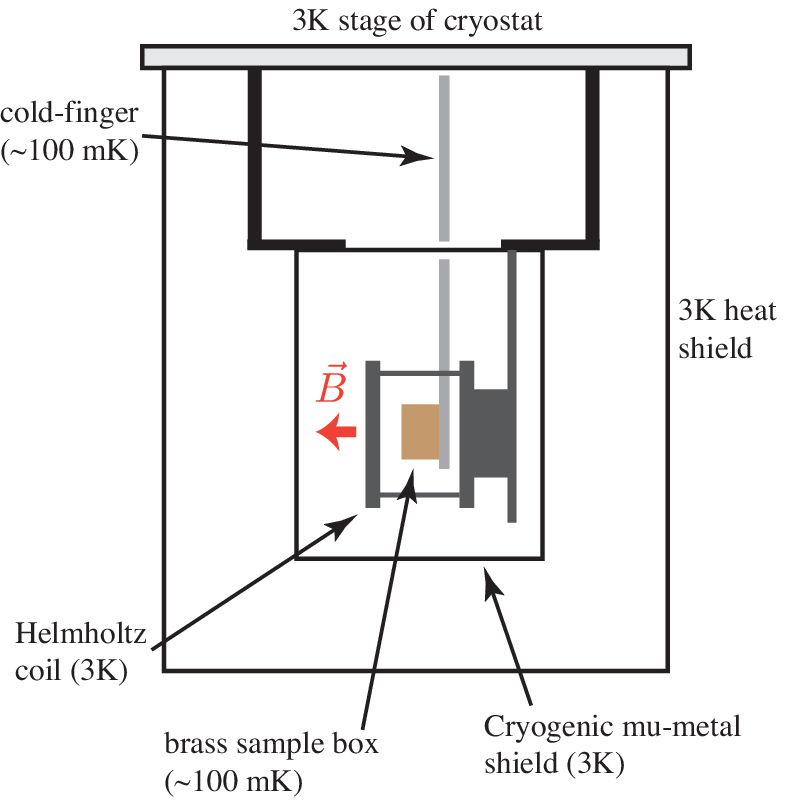}
  \caption{(Color online) Schematic of cryostat including sample box and mount along with Helmholtz coil (not to scale).
  \label{fig:schem}}
\end{figure}

The device is wire-bonded to a microwave board and enclosed in a brass box that is mounted on the cold finger of an adiabatic demagnetization refrigerator (ADR). A $3\,{\rm K}$ pulse-tube cooled stage cools the ADR magnet and also contains a support structure for mounting a superconducting Helmholtz coil and a cryogenic mu-metal can. The cold finger is adjusted so that the sample is positioned at the center of the Helmholtz coil, which is used for applying the cooling fields for trapping vortices. The arrangement of the various components and their associated temperatures is shown in Fig.~\ref{fig:schem}. 
We use a conventional configuration of cold attenuators on the coaxial driveline for exciting the resonance and we amplify the transmission signal with a cryogenic HEMT amplifier on the 3K stage followed by a room-temperature microwave amplifier. 

\section{accounting for current distribution in simulation results}

In order to compare the simulation results in our Letter with the measured internal loss on the harmonic $1/Q_i(B)$, we 
account for the variation of the standing-wave current along the length of the resonator since $n_{qp}(x)$ is proportional to the local effective resistivity. By computing 
$\left(\int_{-L/2}^{L/2} I^2(x) n_{qp}(x) dx\right)/\left(\int_{-L/2}^{L/2} I^2(x) dx\right)$, where $I(x)$ is a full period of a sine wave for the harmonic, and dividing by the $n_{qp}$ value that we extract from zero-field cooling, we can compare this with the measured $1/Q_i(B)$ for the harmonic, normalized by the average of $1/Q_i(B)$ for $B<B_{th}(8\,\mu{\rm m})$.

\section{Variation in loss with cryostat temperature}

\begin{figure}
\centering
\includegraphics[width=3.35in]{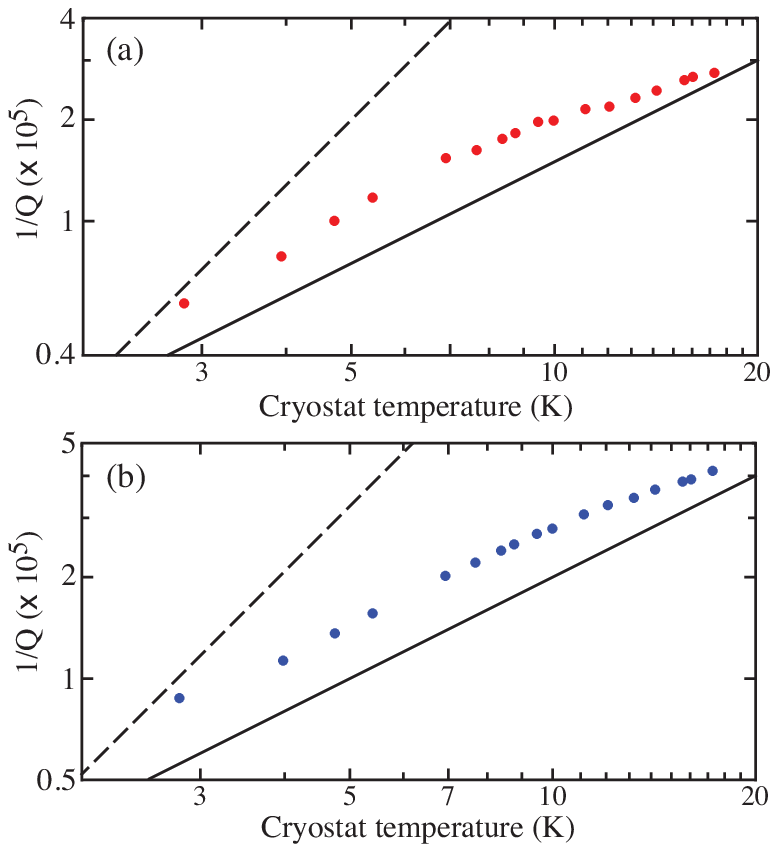}
  \caption{(Color online) Measurements of $1/Q$ vs. cryostat temperature for zero-field cooling for (a) fundamental, (b) harmonic resonance. The temperature of the cold finger and sample remained below $140\,{\rm mK}$ during the measurements. Dashed line is a guide to the eye for a quadratic dependence while the solid line corresponds to a linear dependence.
  \label{fig:T3K}}
\end{figure}

We have performed a similar measurement to Ref.~\cite{Barends:2011s} to confirm the presence of a significant density of nonequilibrium quasiparticles in our resonators due to pair-breaking radiation from warmer parts of the cryostat. By varying the temperature of the cryostat, separate from the cold finger and sample, one can change the radiation power and spectrum that is influencing the resonator. 
For this test we used an identical chip from the same wafer as the one presented in our Letter with the same cryostat configuration. We cooled the resonator with no magnetic field applied with our Helmholtz coil to avoid trapping any vortices. 
By turning off the pulse-tube cooler with the sample at the base temperature, the pulse-tube stage warmed up, thus also warming the Helmholtz coil, magnetic shield, and $3\,{\rm K}$ thermal shield. Even once these components reached $18\,{\rm K}$, the sample temperature increased no higher than $140\,{\rm mK}$. We recorded $S_{21}$ along with the cryostat temperature during this warming process. In Fig.~\ref{fig:T3K} we plot the loss $1/Q$ for the fundamental and harmonic resonance vs. the cryostat temperature. For both resonance modes, the loss increases significantly as the cryostat temperature rises. For a blackbody source with the full spectrum of radiation shining on the resonator, one would expect $1/Q \propto T_{cryostat}^2$ for the arguments put forth in Ref.~\cite{Barends:2011s}. For increased levels of IR shielding surrounding the sample, the radiation spectrum can be cut off, leading to smaller exponents for the increase \cite{Barends:2011s}. Our observed increase in $1/Q$ is closer to linear rather than quadratic, suggesting that our brass sample box that encloses our resonator chip provides some modest filtering of the IR radiation. Nonetheless, the immediate increase in $1/Q$ with $T_{cryostat}$ strongly suggests that nonequilibrium quasiparticles generated by stray IR radiation in our cryostat dominate the loss in our resonator measurements.

\begin{figure}
\centering
\includegraphics[width=3.35in]{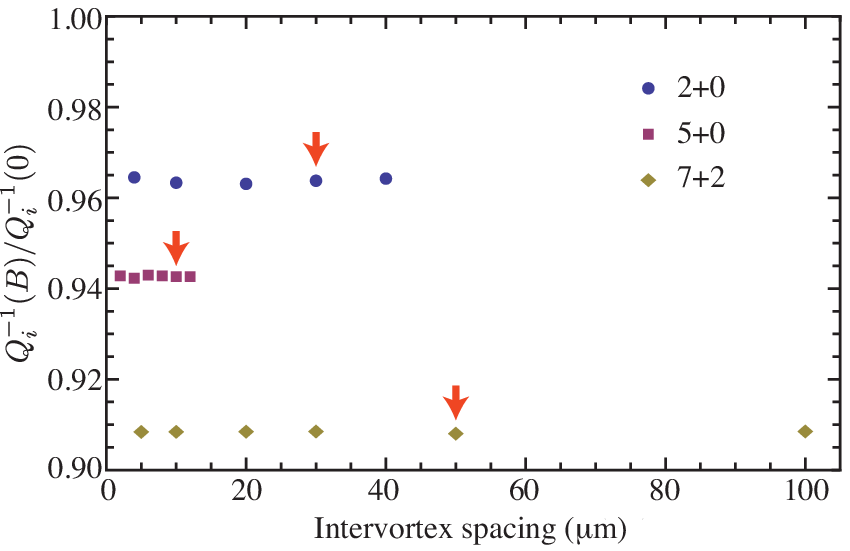}
\caption{(Color online) Simulated normalized quasiparticle loss on harmonic for different intervortex spacings for (i) 2 vortices in central bulge (circles), (ii) 5 vortices in central bulge (squares), (iii) 7 vortices in central bulge that are fixed in place plus 2 vortices in coupling elbow with variable spacing (diamonds). 
Red arrows indicate the intervortex spacing used in the original simulations for the three corresponding points of Fig. 4(b) of the Letter.
\label{fig:sims-vortex-position}}
\end{figure}

\section{Quasiparticle diffusion simulations for different vortex positions}

In our simulations of the quasiparticle diffusion, we include separate vortex-related terms in Eq. (1) in our Letter for each trapped vortex. From the analysis of the step features on the measurements of the fundamental resonance, we can extract the number of vortices in the central bulge region and coupling elbow for different cooling fields. However, we are unable to determine the precise location of each vortex within the bulge or elbow. Thus, for each vortex-number increment in our simulations, we have spaced the vortices evenly within each trapping region. In order to verify that our simulated reduction in quasiparticle density does not depend significantly on the detailed locations of each vortex in the distribution, we have chosen three example steps in the field-dependence from Fig. 4(b) of our Letter: (i) 2 vortices in the bulge and none in the elbow, (ii) 5 vortices in the bulge and none in the elbow, (iii) 7 vortices in the bulge and two in the elbow. For each case we have repeated the simulation for several different values of the intervortex spacing, within the constraints of the size of the bulge and elbow. Figure~\ref{fig:sims-vortex-position} shows the variation in the simulated normalized quasiparticle loss on the harmonic with intervortex spacing for each of these three cases with arrows indicating the spacing values that were used for the corresponding points in Fig. 4(b) of the Letter. There is no significant dependence on the intervortex spacing, thus we conclude that detailed knowledge of the vortex positions in the central bulge and coupling elbow is not necessary for our current modeling of the vortex-quasiparticle interactions.

\section{Simulations with different quasiparticle diffusion constants}

\begin{figure}
\centering
\includegraphics[width=3.35in]{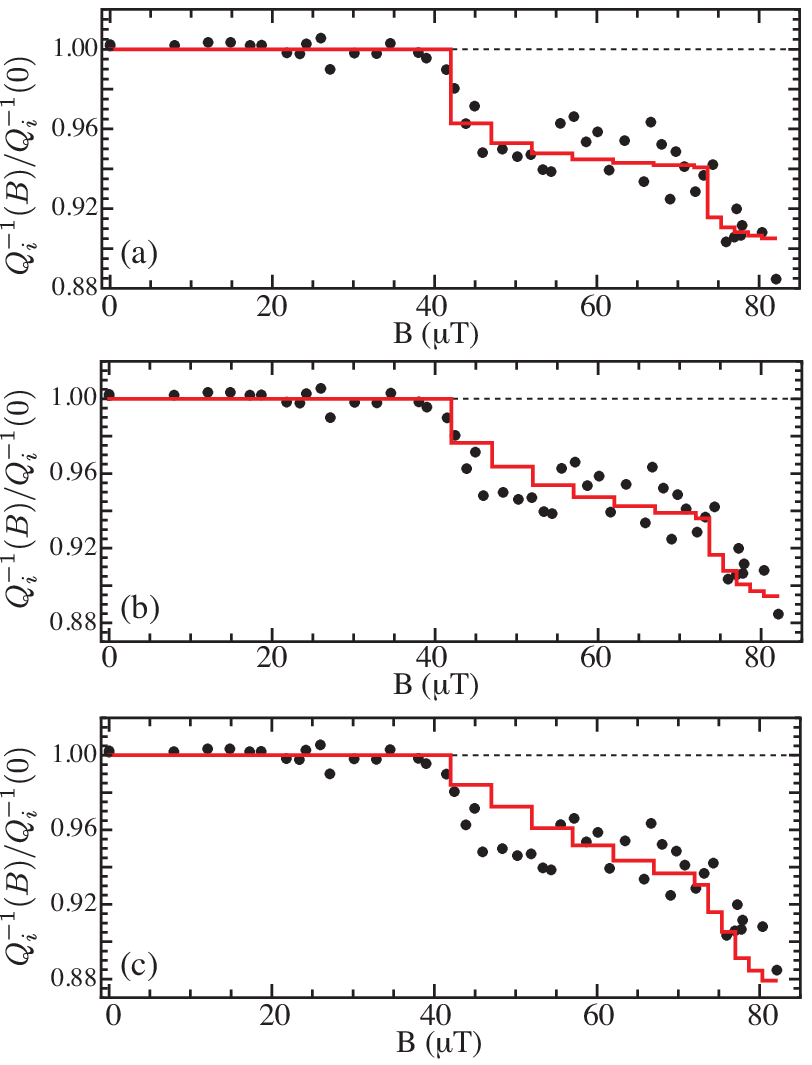}
\caption{(Color online) Measured $1/Q_i(B)$ for harmonic, normalized by average of $1/Q_i$ below threshold field (points); simulations of normalized quasiparticle loss on harmonic for different parameters (solid line): (a) $D=30\,{\rm cm^2/s}$, $\Gamma_R=20\,\mu{\rm m}^3/s$, $\Gamma_v=7\times10^6\,{\rm s}^{-1}$; (b) $D=60\,{\rm cm^2/s}$, $\Gamma_R=30\,\mu{\rm m}^3/s$, $\Gamma_v=3.5\times10^6\,{\rm s}^{-1}$; (c) $D=150\,{\rm cm^2/s}$, $\Gamma_R=40\,\mu{\rm m}^3/s$, $\Gamma_v=2\times10^6\,{\rm s}^{-1}$. Note: the plot in (b) is identical to Fig. 4(b) in the Letter and is repeated here for comparison with the other simulations.
\label{fig:sims-diffusion}}
\end{figure}

There have been many investigations of quasiparticle dynamics in Al films at low temperatures and the effective quasiparticle diffusion constants $D$ reported in the literature can be influenced by multiple factors. The 
normal metal diffusion constant $D_n$ directly affects $D$ and there is a range of reported values of $D_n$ for Al films, including $49\,{\rm cm^2/s}$ \cite{Anthore:2003s} and $140\,{\rm cm^2/s}$ \cite{Pekola:2000s}. Of course, such variations can be caused by different electronic mean free paths depending on the film quality in the various experiments. Also, $D$ will be reduced from $D_n$ depending on the quasiparticle energy: $D(E) = D_n (1-(\Delta/E)^2)^{1/2}$ \cite{Ullom:1998s}, as we described in our Letter. However, even after accounting for the reduction in $D$ due to the quasiparticle energy, there is evidence that the effective $D$ is typically reduced further still \cite{Friedrich:1997s}. For our Al film, we estimate $D_n = 150\,{\rm cm^2/s}$ based on the measured resistivity at $4\,{\rm K}$ of $0.5\,\mu\Omega$-cm. In order to account for the anomalous reduction described in Ref.~\cite{Friedrich:1997s}, we used $D_n = 60\,{\rm cm^2/s}$, combined with an estimate for the approximate quasiparticle energy as described in the Letter, to determine $D$ for the simulations presented in Fig.~4(b). We then explored the sensitivity of our model to the value of $D$ used in the simulations by running our simulations from Fig.~4(b) for two other values of $D_n$: $30$ and $150\,{\rm cm^2/s}$ (Fig.~\ref{fig:sims-diffusion}). In each case, we adjusted the values of $\Gamma_R$ and $\Gamma_v$ to give the best agreement between the simulated curve and the normalized measured loss vs. field for the harmonic; the resulting values are listed in the caption to Fig.~\ref{fig:sims-diffusion}. For smaller $D_n$, we obtained the best match to the data for smaller $\Gamma_R$ and larger $\Gamma_v$. The resulting values for $\Gamma_R$ for all three cases are well within the range reported by others for Al films \cite{Ullom:1998s} and all three resulting $\Gamma_v$ values are also quite consistent with typical electron-phonon scattering rates for Al thin films at low temperatures \cite{Kaplan:1976s,Chi:1979s}. Thus, our model of the quasiparticle diffusion and interaction with vortices is able to provide a reasonable description of our experimental measurements over a range of parameters for quasiparticle dynamics, consistent with the variation in values for quasiparticle dynamics in Al films reported in the literature.

\section{Threshold cooling fields for vortex trapping}
\begin{figure}
\centering
\includegraphics[width=3.35in]{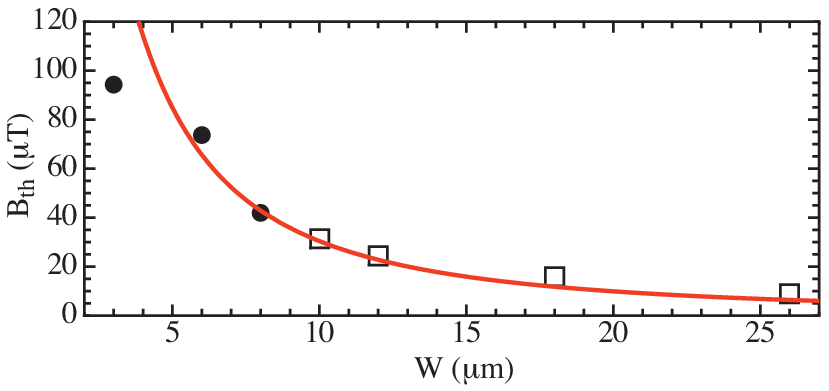}
\caption{(Color online) Plot of threshold cooling fields $B_{th}$ for different width segments on resonator from Letter (filled circles) and $B_{th}(w)$ values for quarter-wave uniform-width resonators from separate device as discussed in text (open squares). Curve corresponds to Eq.~(\ref{eq:threshold}) for $\xi=250\,{\rm nm}$. 
\label{fig:threshold-fields}}
\end{figure}
The relationship between the width of a superconducting strip and the value of $B_{th}$ was studied in Ref.~\cite{Stan:2004s} with field-cooling followed by imaging of the vortex distributions. The extracted values of $B_{th}$ for strips of different width $w$ were found to be in reasonable agreement with the expression 
\begin{equation}
B_{th}=\frac{2 \Phi_0}{\pi w^2} \ln\left( \frac{w}{4\xi} \right),
\label{eq:threshold}
\end{equation}
where $\Phi_0 \equiv h/2e$ is the superconducting flux quantum and $\xi$ is the coherence length at the temperature at which the vortices freeze into their respective pinning sites \cite{Stan:2004s,likharev72s}. From our measurements of $1/Q_v(B)$ in our Letter, we have extracted values of $B_{th}$ for $w=3, 6, 8\,\mu{\rm m}$ for the three characteristic widths in the different regions of our resonator and we plot these values in Fig.~\ref{fig:threshold-fields}. Because this is a rather narrow range of $w$ to compare with Eq.~(\ref{eq:threshold}), we have chosen to include some previously unpublished $B_{th}$ data from our lab on some other Al resonators with a different geometry, but a wider range of linewidths. This other chip contained four quarter-wave coplanar waveguide resonators with uniform-width center conductors, similar to the device in Ref.~\cite{Song:2009s}, with widths $w=10, 12, 18, 26\,\mu{\rm m}$. Also, the thickness of the Al film on this other chip was $150\,{\rm nm}$, although it is not clear what role, if any, film thickness plays in determining $B_{th}$. By analyzing the $1/Q_v$ measurements for this chip, we have extracted $B_{th}$ for the four resonators of different widths and we include this data in Fig.~\ref{fig:threshold-fields} with the $B_{th}(w)$ points extracted from the measurements in our Letter. We then include a curve corresponding to Eq.~(\ref{eq:threshold}) by adjusting $\xi$. We find that for $\xi=250\,{\rm nm}$, we obtain decent agreement with the measured $B_{th}(w)$ points, although the curve is not a perfect match to the data. Deviations between the measurements and the predicted dependence of Eq.~(\ref{eq:threshold}) could be due to a variety of reasons, as discussed in Ref.~\cite{Stan:2004s}, such as variations in the details of the vortex freezing process between the strips of different widths. Also, for some of our features, such as the $6\,\mu{\rm m}$ and $8\,\mu{\rm m}$ regions of our resonator, the finite length of these regions may change the details of Eq.~(\ref{eq:threshold}) as well. Nonetheless, the general trend of $B_{th}$ is clear and vortices trap at higher threshold fields for narrower superconducting traces. Thus, our  scheme for making resonators with variable widths of the center conductor allows for the control of vortex-trapping locations along the resonator length. 


\end{document}